\documentstyle[12pt,epsf]{article}

\begin{document}

\title{Market price simulator based on analog electrical circuit}
\author{Aki-Hiro Sato$^{1}$\footnote{Electronic
mail:aki@sawada.riec.tohoku.ac.jp} and Hideki Takayasu$^{2}$ \\
Department of Applied Mathematics and Physics, \\
Kyoto University, Kyoto 606-8501, Japan, and \\
$^{2}$ Sony Computer Science Lab., Takanawa Muse Bldg., 3-14-13, \\
Higashi-Gotanda, Shinagawa-ku, Tokyo 141-0022, Japan.
}

\maketitle              

\begin{center}
{\bf Abstract}
\end{center}
{
We constructed an analog electrical circuit which generates fluctuations in 
which probability density function has power law tails. In the
circuit fluctuations with an arbitrary exponent of the power law can be 
obtained by adjusting the resistance. With this low cost circuit the 
random fluctuations which have the similar statistics to foreign exchange
rates can be generated as fast as an expensive digital computer.
}

\section{Introduction}
\label{sec:introduction}
Random multiplicative process (RMP) is attracting researchers
as a new mechanism of generating fat tails in the distribution
of price changes in open markets recently (Takayasu et al. 1997, 
J\"ogi et al. 1998, Sato et al. 2000). One of the mechanisms of the
fat tails is a feedback mechanism of both positive and negative
in a random manner. The positive feedback plays a role of
amplification while the negative one plays a role of damping. In the
presence of positive feedback one may consider that the system is
unstable, however, this is not always correct. When the damping effects
dominates the amplification the system can generate statistically
stationary fluctuations. 

Recently fat tails in the distribution have been confirmed in the market price 
fluctuation (Mantegna et al. 1995). Among several approaches to this 
phenomena (Bak et al. 1997, Lux et al. 1999), the authors demonstrated
that RMP can be derived from a microscopic artificial dealer's
market simulation model (Sato et al. 1998). The RMP has also been
derived from a macroscopic theoretical analysis of the critical
dynamics of the balance of demand and supply (Takayasu et al. 1999). 
The multiplicative effect of the market comes from the dealer's forecasting 
which follows the trend of the latest market price changes, which 
obviously enhances the fluctuations in price.

\begin{figure}[h]
\begin{center}
\epsfxsize=220pt
\epsfbox{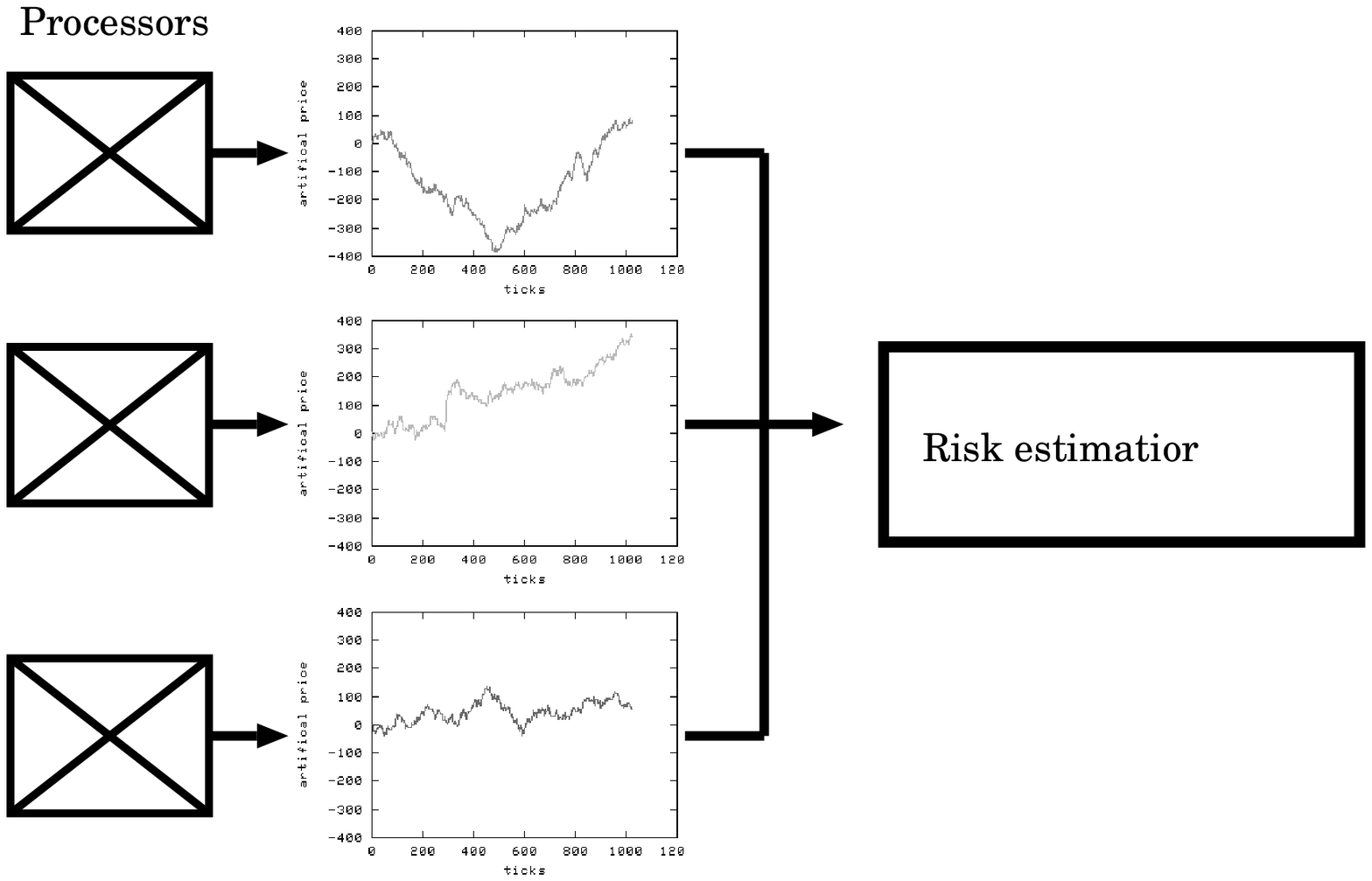}
\end{center}
\caption{Conceptual illustration of risk estimator that calculates 
realizations of market prices directly by parallel processing units.}
\label{fig:parallel}
\end{figure}

In this article we review the analog electrical circuit which generates 
power law fluctuations based on the RMP theory. The application of
this circuit is to generate fluctuation which is statistically similar to 
market price change in inexpensive way at high speed. The goal of this 
study is a risk estimator that calculates realizations of market prices 
directly by parallel processing units and calculates a statistical 
feature of the fluctuation from these simulated data in the meaning of 
an ensemble (see. fig. \ref{fig:parallel}.).  

The outline of this article is the followings. In sec. 
\ref{sec:foreign-currency-rate}
we analyze tick data of yen/dollar exchange rates, and show that a 
probability density function of the market price changes has fat
tails of roughly power law. 
In sec. \ref{sec:theory} a random multiplicative process is 
formalized by a stochastic differential 
equation with a multiplicative noise. It is indicated that the 
probability density of a dynamical variable follows 
a power law distribution. In sec. \ref{sec:circuit} an example of electrical
circuit diagram is shown and an equivalent equation of the proposed circuit 
is described. In sec. \ref{sec:observation} we show the result of 
observation and we will discuss them. 
Sec. \ref{sec:conclusions} is devoted to the concluding remarks.

\section{Statistical properties of foreign currency rate}
\label{sec:foreign-currency-rate}
\begin{figure}[h]
\begin{center}
\epsfxsize=220pt
\epsfbox{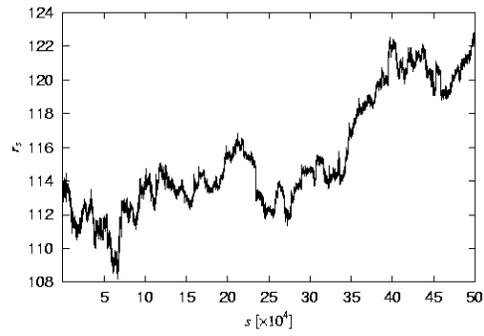}
\epsfxsize=220pt
\epsfbox{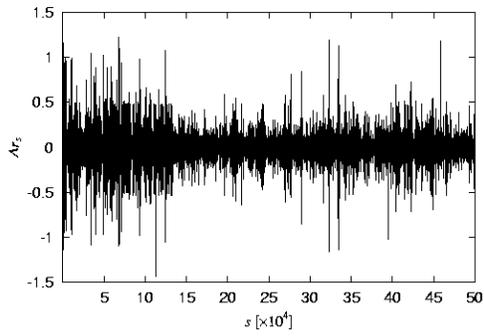}
\end{center}
\caption{Typical examples of yen/dollar exchange rates (the above) and 
its changes (the bellow) from January 1999 to March 1999.}
\label{fig:yen-dollar-seq}
\end{figure}

\begin{figure}[h]
\begin{center}
\epsfxsize=220pt
\epsfbox{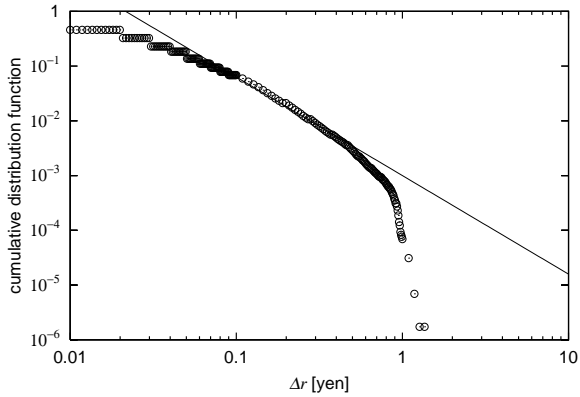}
\end{center}
\caption{The log-log plots of the cumulative distribution function of 
yen/dollar exchange rates. A solid line represents the power law with 
$\beta = 1.8$.} 
\label{fig:yen-dollar-cdf}
\end{figure}

Mantegna et al. investigated time series of stock market index
of S\&P500 and reported in their famous paper (Mantegna et al. 1995) 
that a probability density function of changes obeys a power law 
distribution. They indicated that the power law exponent is estimated 
as 1.4. However, it is clarified that the exponent is not
universal by other researches. Besides stock market prices 
it is known that the probability density function of foreign exchange 
rates also follows a power law distribution (Takayasu et al. 2000).
We analyze tick data of yen/dollar exchange rate of 578,508 data points
from January to March 1999. Let us denote the rate of the $s$th tick as $r_s$. 
The change of the rate is defined as $\Delta r_s = r_s - r_{s-1}$. 
Fig. \ref{fig:yen-dollar-seq} shows typical examples of time series 
of yen-dollar exchange rates and corresponding changes. The cumulative 
distribution function (CDF), which is defined by
\begin{equation}
F(\geq |x|) = \int_{-\infty}^{-|x|} p(x')dx' +
\int_{|x|}^{\infty}p(x')dx',
\label{eq:cumulative}
\end{equation}
is shown in fig. \ref{fig:yen-dollar-cdf}. 
It is clear that the CDF has a liner slope with a quick decay
by 1 yen in log-log plots. From the liner slope in log-log plots 
we can estimated the exponent of the power law distribution,
\begin{equation}
F(\geq |x|) \propto |x|^{-\beta}.
\end{equation}
The best fit value of $\beta$ we obtain from the exchange rate is 
$\beta=1.8$.

\section{Theory}
\label{sec:theory}
In this section we show a brief outline of a continuous time version 
of random multiplicative process. The random multiplicative process is 
given by a stochastic differential equation,
\begin{equation}
\frac{dv}{dt} = \nu(t)v(t) + \xi(t),
\label{eq:RMP}
\end{equation}
where $v(t)$ is a dynamic variable and $\nu(t)$ and $\xi(t)$ represent 
a multiplicative noise and an additive 
noise, respectively. Denoting an ensemble average by $\langle \ldots \rangle$, 
we assume the following relations for both multiplicative and additive noises 
based on the Gaussian white noise theory.
\begin{eqnarray}
\langle \nu(t) \rangle &=& \bar{\nu}, \\
\langle [\nu(t_1) - \bar{\nu}][\nu(t_2) - \bar{\nu}] \rangle &=& 2 D_{\nu}
\delta(t_1 - t_2), \\
\langle \xi(t) \rangle &=& 0, \\
\langle \xi(t_1)\xi(t_2) \rangle &=& 2 D_{\xi}\delta(t_1-t_2),
\end{eqnarray}
where $\bar{\nu}$ represents an average of the multiplicative noise, and 
$D_{\nu}$ and $D_{\xi}$ represent the strength of the multiplicative 
noise and that of the additive noise, respectively.

The probability density function of $v(t)$  in eq. (\ref{eq:RMP}) is
denoted as $p(v,t)$, is known to follow the generalized Fokker-Planck 
equation (Deutsch 1994, Venkataramani et al. 1996, Nakao 1998).
\begin{equation}
\frac{\partial}{\partial t}p(v,t) = \frac{\partial}{\partial
v}\Bigl[-(\bar{\nu}+ D_{\nu})v p(v,t) + \frac{\partial}{\partial
v}(D_{\nu}v^2 + D_{\xi})p(v,t)\Bigr]. 
\label{eq:Fokker-Planck}
\end{equation}
By solving eq. (\ref{eq:Fokker-Planck}) for a steady state 
$\frac{\partial}{\partial t}p(v,t)=0$ and boundary condition,
$\frac{\partial}{\partial v}p(\pm v_u,t)=0$, we get the stationary 
distribution.
\begin{equation}
p(v) \propto (D_{\xi}+D_{\nu}v^2)^{\frac{\bar{\nu}}{2D_{\nu}}-\frac{1}{2}},
\label{eq:steady-state-pdf}
\end{equation}
which has power law tails for large $v$, 
\begin{equation}
p(v) \propto |v|^{-\beta-1},
\label{eq:define-of-power-law-exponent}
\end{equation}
where $\beta = -\frac{\bar{\nu}}{D_{\nu}}$. In other words the power law 
exponent is given by a simple function of the average and variance of the
multiplicative random noise.

\section{The Circuit}
\label{sec:circuit}
\begin{figure}[h]
\begin{center}
\epsfxsize=220pt
\epsfbox{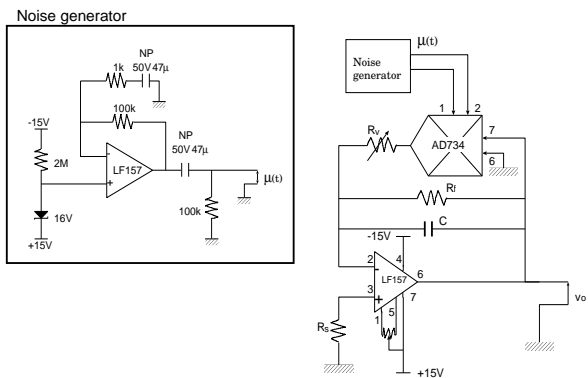}
\end{center}
\caption{Block diagram of the circuit with a noise generator. The
circuit contains a noise generator, a analog multiplier and
an operational amplifier for integration. $R_f=100$k$\Omega$,
$C=10$pF,$R_v=200\Omega$. A variable resistor under the operational 
amplifier is for adjustment of the offset.}
\label{fig:circuit}
\end{figure}

From the assumption for the multiplicative noise in the RMP theory 
described in sec. \ref{sec:theory},
it is important that $\nu(t)$ takes both positive and negative values
to realize the power law tails. With an electrical analog circuit this 
means that it is necessary for the circuit to include both positive and 
negative feedbacks in the meaning of probability. We solve this
problem by using an analog multiplier. Our block diagram of 
circuit and an implemented noise generator are shown in 
fig. \ref{fig:circuit}. $v_o$ in the figure represents an output voltage, 
and $\mu(t)$ is an output directly from 
the noise generator. As it is seen from this figure the circuit 
contains the noise generator, an analog 
multiplier (Analog devices, 10MHz, 4-quadrant)
and an operational amplifier (National Semiconductor, LF157) for
integrator (Sato et al. 2000). 

The output of the noise generator plays a role of the multiplicative
noise in eq. (\ref{eq:RMP}). In the noise generator a shot noise
of the zener diode in fig. \ref{fig:circuit} is amplified by an
operational amplifier, and it is output from generator through high pass
filter. LF157 has the 20M product of a voltage gain (G) 
and a bandwidth (B). The bandwidth is given by B=200kHz in the noise 
generator because we put G=100. Thus we expect a frequency
characteristics of the noise generator to be up to 200kHz . 
Let us explain how to realize both positive and negative 
feedbacks in the meaning of probability. We realize both positive and negative 
feedbacks by multiplying $v_o$ with the output of the noise generator 
$\mu(t)$ in the analog multiplier 
and by connecting the product to the negative input of the operational
amplifier for integrator. Although the additive noise term, $\xi(t)$, is 
not explicitly added in the circuit, it derives from either thermal noises 
of the operational amplifier or from an external electro-magnetic noises.

An equation equivalent to the circuit diagram of fig. \ref{fig:circuit} 
is given as
\begin{equation}
\frac{dv_o}{dt} 
= \Bigl(\frac{1}{R_{f}C}+\frac{k}{R_{v}C}\mu(t)\Bigr)v_o + \xi(t),
\label{eq:circuit}
\end{equation}
where $k$ is a factor of the multiplier and $k=1/10$.
$\xi(t)$ represents the additive noise effect. The strength of multiplicative
noise $\mu(t)$ depends on the value of a variable resistor $R_v$ in
fig. \ref{fig:circuit} because $R_v$ is a factor of $\mu(t)$ in
eq. (\ref{eq:circuit}). Therefore, we expect that the power law exponent for
the output is  a function of $R_v$. 

\section{Results of observation and discussion}
\label{sec:observation}
We measure the output of the circuit $v_o$ for $R_v$ since 
we expect that the power law exponent $\beta$ is a function of $R_v$ 
from the above discussion. We obtained the output through a 12-bit AD
converter (Microscience, ADM-652AT) and processed it as digital data in
digital computer. A sampling frequency is 125kHz throughout all the 
observations.

\begin{figure}[h]
\begin{center}
\epsfxsize=220pt
\epsfbox{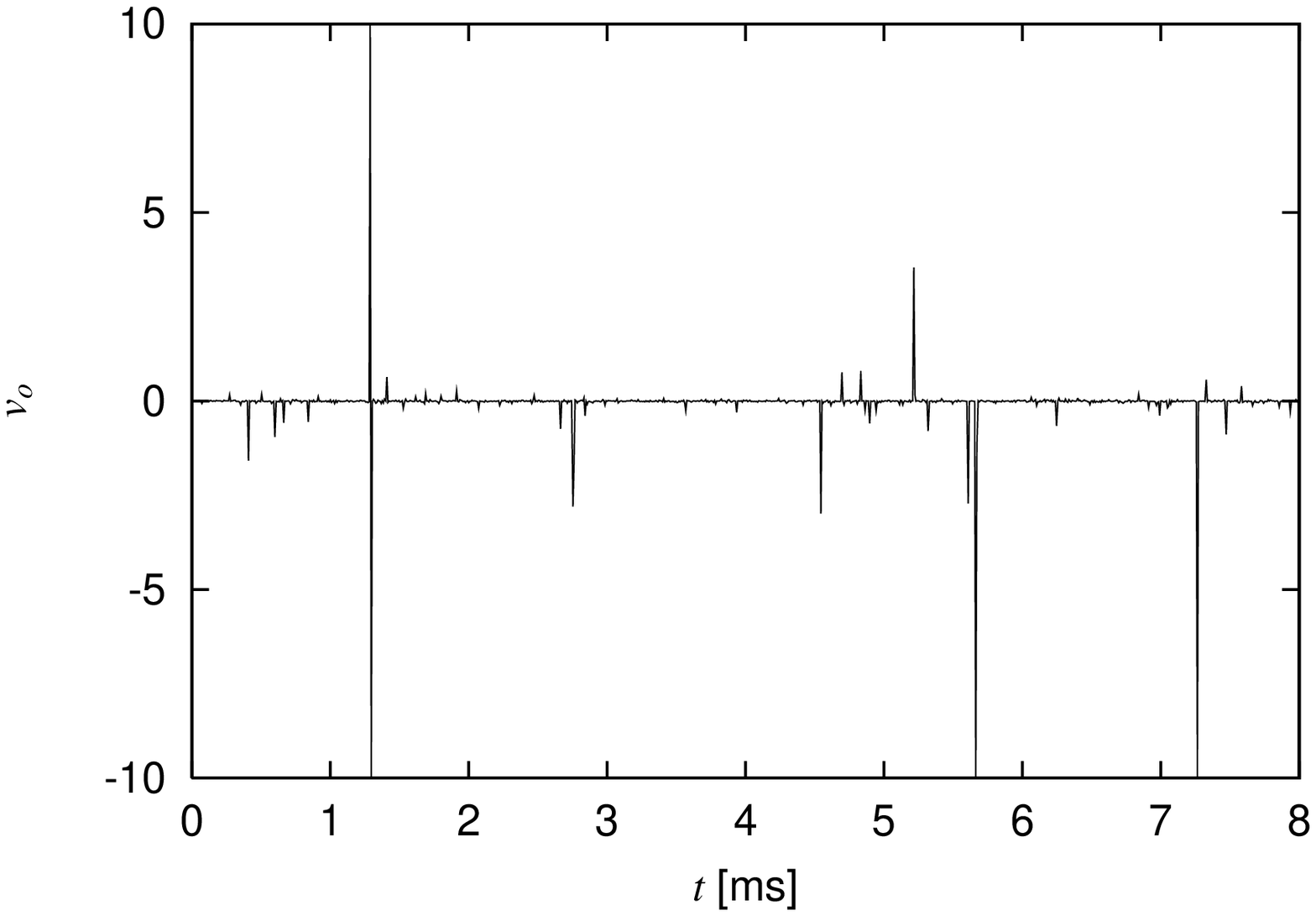}
\epsfxsize=220pt
\epsfbox{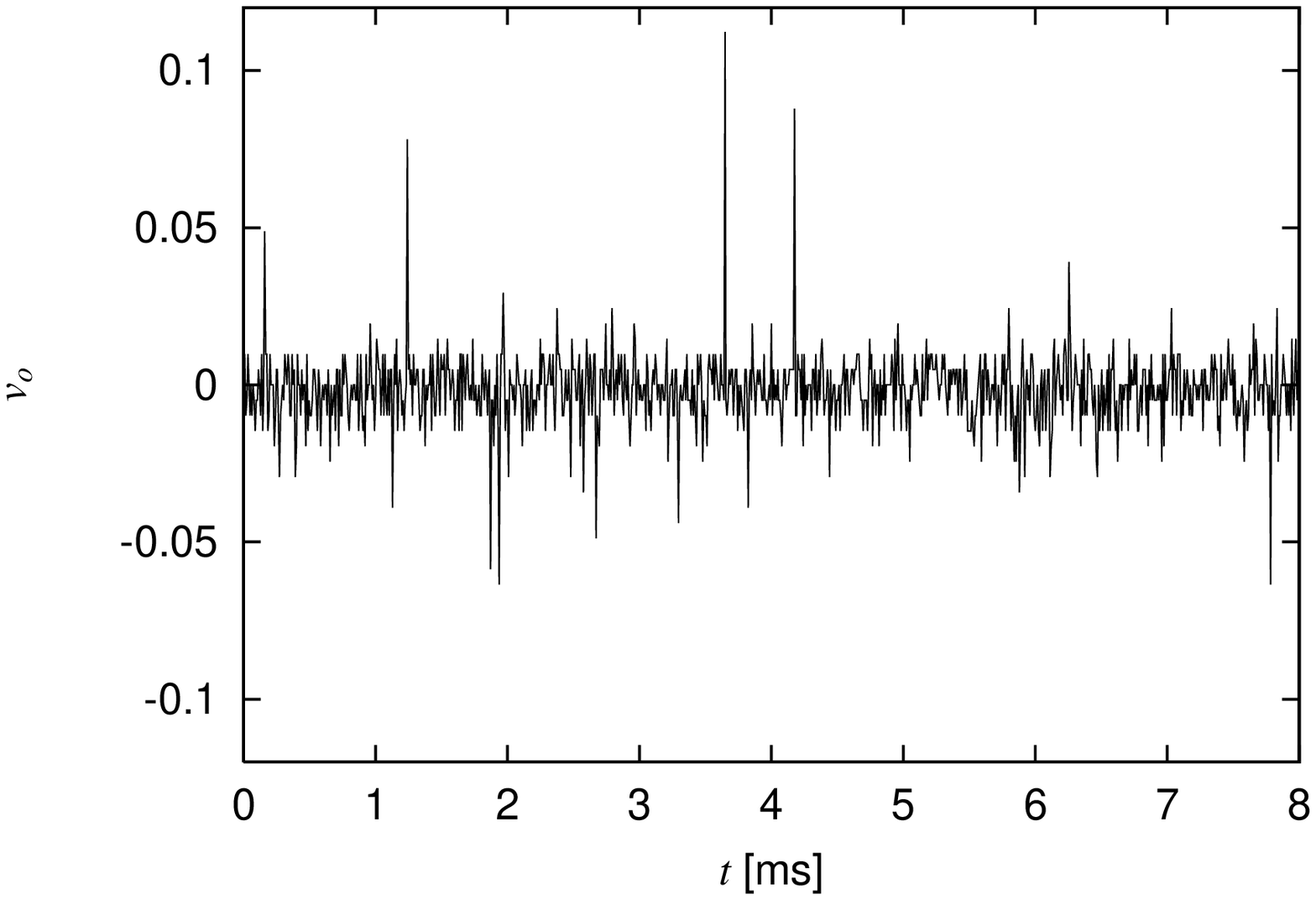}
\end{center}
\caption{Typical examples of time series at $R_v=50\Omega$ and $100\Omega$.}
\label{fig:seqs}
\end{figure}

\begin{figure}[h]
\begin{center}
\epsfxsize=220pt
\epsfbox{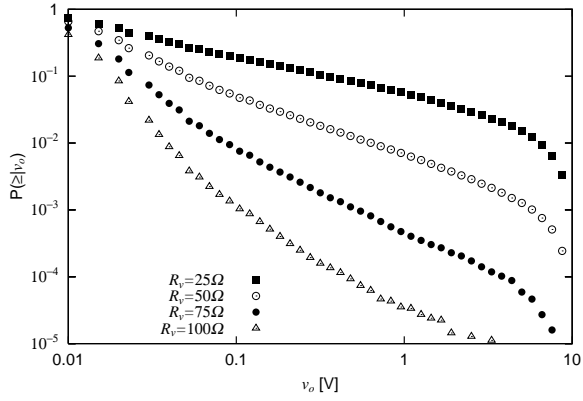}
\end{center}
\caption{Log-log plots of CDF of the output $v_o(t)$ at 
$R_v=25,50,75,100\Omega$. Each CDF shows straight lines with 
different slopes between a decade.}
\label{fig:cdfs}
\end{figure}

\begin{figure}[h]
\begin{center}
\epsfxsize=220pt
\epsfbox{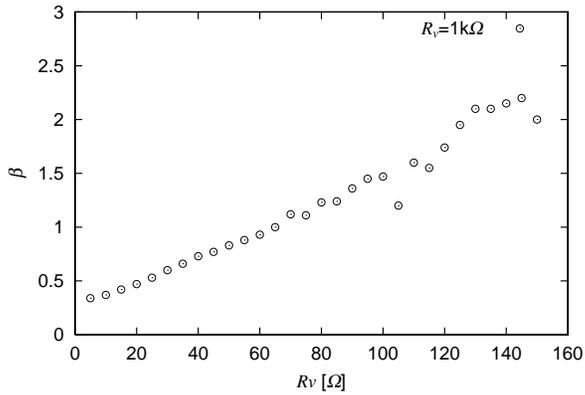}
\end{center}
\caption{The power law exponent $\beta$ plotted against values of a
variable resistance $R_v$. The power law exponent $\beta$ is
numerically estimated from a slope of a liner part in log-log plots of CDF.}
\label{fig:beta}
\end{figure}

\begin{figure}[h]
\begin{center}
\epsfxsize=220pt
\epsfbox{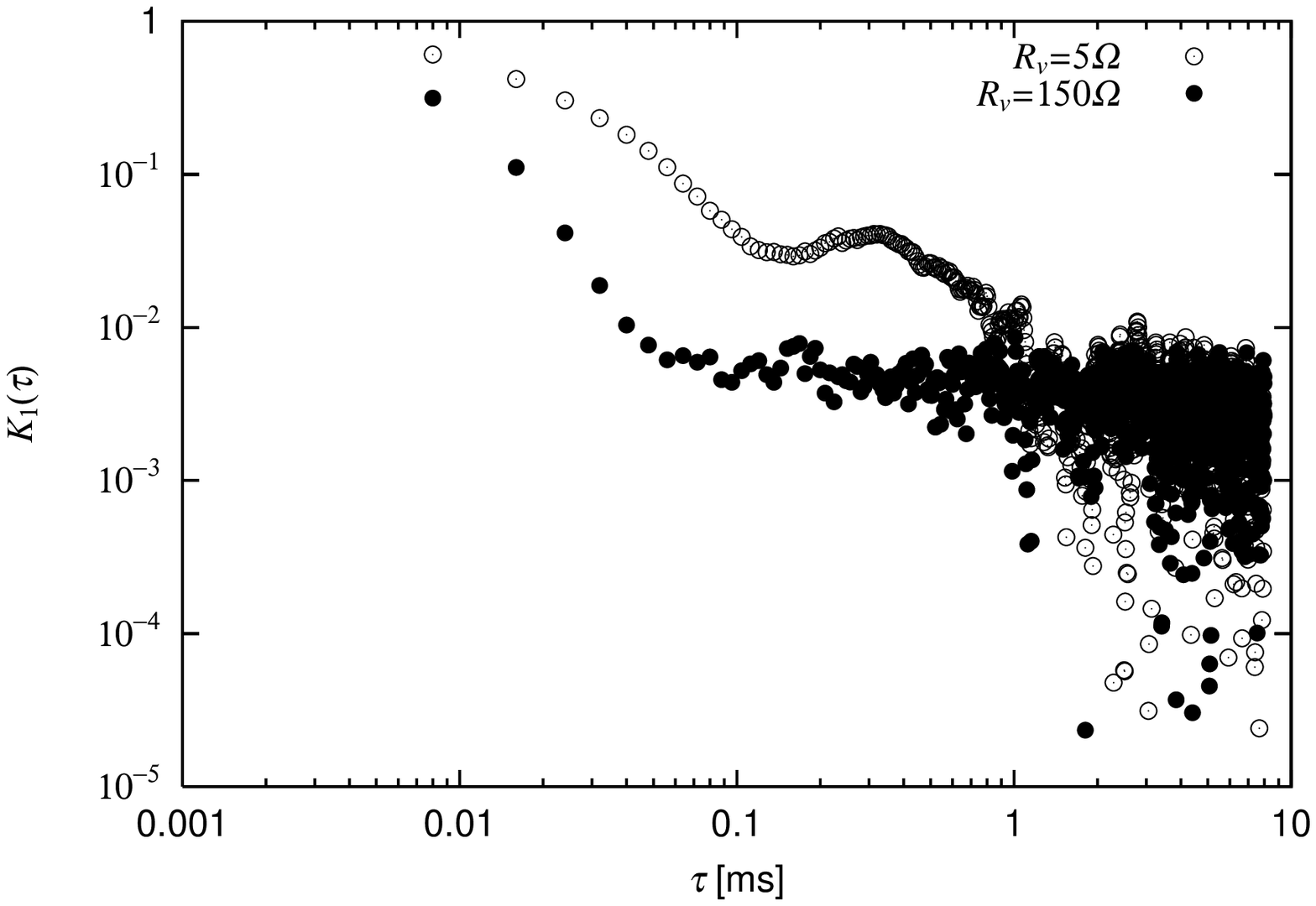}
\epsfxsize=220pt
\epsfbox{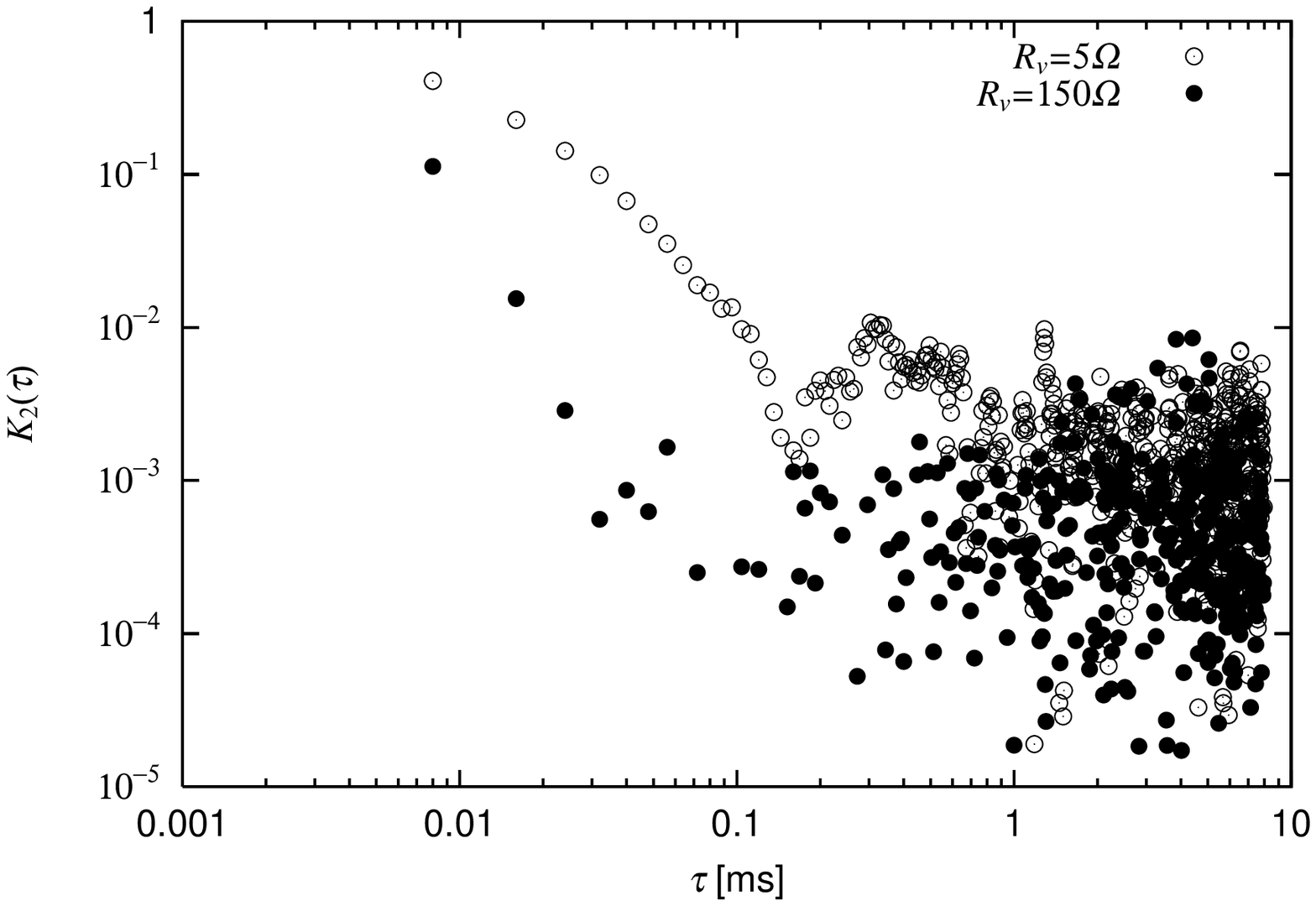}
\end{center}
\caption{Autocorrelation function and volatility autocorrelation
function of output $v_o(t)$.}
\label{fig:acorr}
\end{figure}

We show a typical example of time series of $v_o$ at $R_v=50\Omega$ and 
$100\Omega$ in fig. \ref{fig:seqs}. The time series at $R_v=50\Omega$
sometimes exhibits larger fluctuations than $R_v=100\Omega$. We show 
log-log plots of CDF of $v(t)$ at $R_v=25$,$50$,$75$,$100$$\Omega$ in 
fig. \ref{fig:cdfs}. 
Log-log plots of the CDF have liner parts for about a decade. We clearly find 
that the exponent $\beta$ depends on $R_v$ as expected
qualitatively. Moreover as shown in fig. \ref{fig:beta} we find a
linear relation between $\beta$ and $R_v$. The autocorrelation 
function $R(\tau)$ and the volatility autocorrelation 
function $R^{(2)}(\tau)$ are defined by 
\begin{eqnarray}
R(\tau) &=& \langle v_o(t+\tau)v_o(t) \rangle - \langle v_o(t+\tau) \rangle \langle v_o(t) \rangle, \\
R^{(2)}(\tau) &=& \langle v_o(t+\tau)^2 v_o(t)^2 \rangle - \langle v_o(t+\tau)^2 \rangle \langle v_o^2(t) \rangle.
\end{eqnarray}
We show the autocorrelation function and volatility autocorrelation
function in fig. \ref{fig:acorr}. Both autocorrelation function
and volatility autocorrelation function have quick decay. 
It is known that the volatility autocorrelation function of real market
price changes has a long time correlation. The disagreement of the
volatility autocorrelation occurs because the model 
equation is too simple to describe time structure of fluctuations. 
We need to study higher order differential equations with multiplicative 
noises.

\section{Conclusions}
\label{sec:conclusions}
We proposed an analog electrical circuit as an analog generator of power 
law fluctuations. We described a theoretical equation representing
the electrical circuit and showed that the probability density function of 
a dynamical variable has power law tails. We measured the output of the 
circuit and observed its cumulative distribution functions. The cumulative 
distribution of output voltage has power law tails. The power law exponent 
can be tuned by controlling the variable resistance $R_v$.
We expect that  the proposed circuit is applicable to generate 
fluctuations having power law distribution in a
much cheaper way than any digital computing methods. Moreover,
fluctuations of the circuit may be of use for risk estimation in
foreign exchange or stock market in the near future.

\vspace{2ex}

\noindent
{\bf Acknowledgment}

One of the authors (A.-H. Sato) thanks to Yoshihiro Hayakawa for stimulative
discussion. This research is partially supported by Japan Society for 
the Promotion of Science.


\end{document}